\documentclass[aps,prb,twocolumn,groupedaddress, showpacs]{revtex4-1}
\usepackage{graphicx,amssymb,bm,natbib,amsfonts,amsmath,epsfig,epstopdf}
\setcitestyle{super}
\newcommand*{\citen}[1]{%
  \begingroup
    \romannumeral-`\x 
    \setcitestyle{numbers}%
    \cite{#1}%
  \endgroup
}

\begin{document}
\title{High field response of gated graphene at THz frequencies}

\author{Hadi Razavipour}
\affiliation{Department of Physics, McGill University, Montr\'eal, Qu\'ebec, Canada}
\author{Ibraheem Al-Naib}
\affiliation{Department of Physics, Engineering Physics and Astronomy, Queen's University, Kingston, Ontario, Canada}
\author{Wayne Yang}
\affiliation{Department of Physics, McGill University, Montr\'eal, Qu\'ebec, Canada}
\author{Fran\c{c}ois Blanchard}
\affiliation{D\'epartement de g\'enie \'Electrique, \'Ecole de technologie sup\'erieure, Montr\'eal, Qu\'ebec, Canada}
\author{Abdeladim Guermoune}
\affiliation{Department of Physics, McGill University, Montr\'eal, Qu\'ebec, Canada}
\author{Michael Hilke}
\affiliation{Department of Physics, McGill University, Montr\'eal, Qu\'ebec, Canada}
\author{Marc M. Dignam}
\affiliation{Department of Physics, Engineering Physics and Astronomy, Queen's University, Kingston, Ontario, Canada}
\author{David G. Cooke$^*$}
\affiliation{Department of Physics, McGill University, Montr\'eal, Qu\'ebec, Canada}
\email{cooke@physics.mcgill.ca}
\author{Hassan A. Hafez, Xin Chai, Denis Ferachou, and Tsuneyuki Ozaki}
\affiliation{INRS-EMT, Advanced Laser Light Source, Varennes, Qu\'ebec, Canada}
\author{Pierre L\'{e}vesque and Richard Martel}
\affiliation{D\'{e}partement de Chimie, Universit\'{e} de Montr\'{e}al, Montr\'{e}al, Qu\'ebec, Canada}

\begin{abstract}
We study the Fermi energy level dependence of nonlinear terahertz (THz) transmission of gated multi-layer and single-layer graphene transferred onto sapphire and quartz substrates. The two samples represent two limits of low-field impurity scattering: short-range neutral and long-range charged impurity scattering, respectively.  We observe an increase in the transmission as the field amplitude is increased due to intraband absorption bleaching starting at fields above 8 kV/cm. This effect arises from a field-induced reduction in THz conductivity that depends strongly on the Fermi energy.  We account for intraband absorption using a free carrier Drude model that includes neutral and charged impurity scattering as well as optical phonon scattering.  We find that although the Fermi-level dependence in the monolayer and five-layer samples is quite different, both exhibit a strong dependence on the field amplitude that cannot be explained on the basis of an increase in the lattice temperature alone.  Our results provide a deeper understanding of transport in graphene devices operating at THz frequencies and in modest kV/cm field strengths where nonlinearities exist. 
\end{abstract}
\maketitle
\section{Introduction}
\begin{figure}[htp]
    \centering
    \includegraphics[width=8 cm]{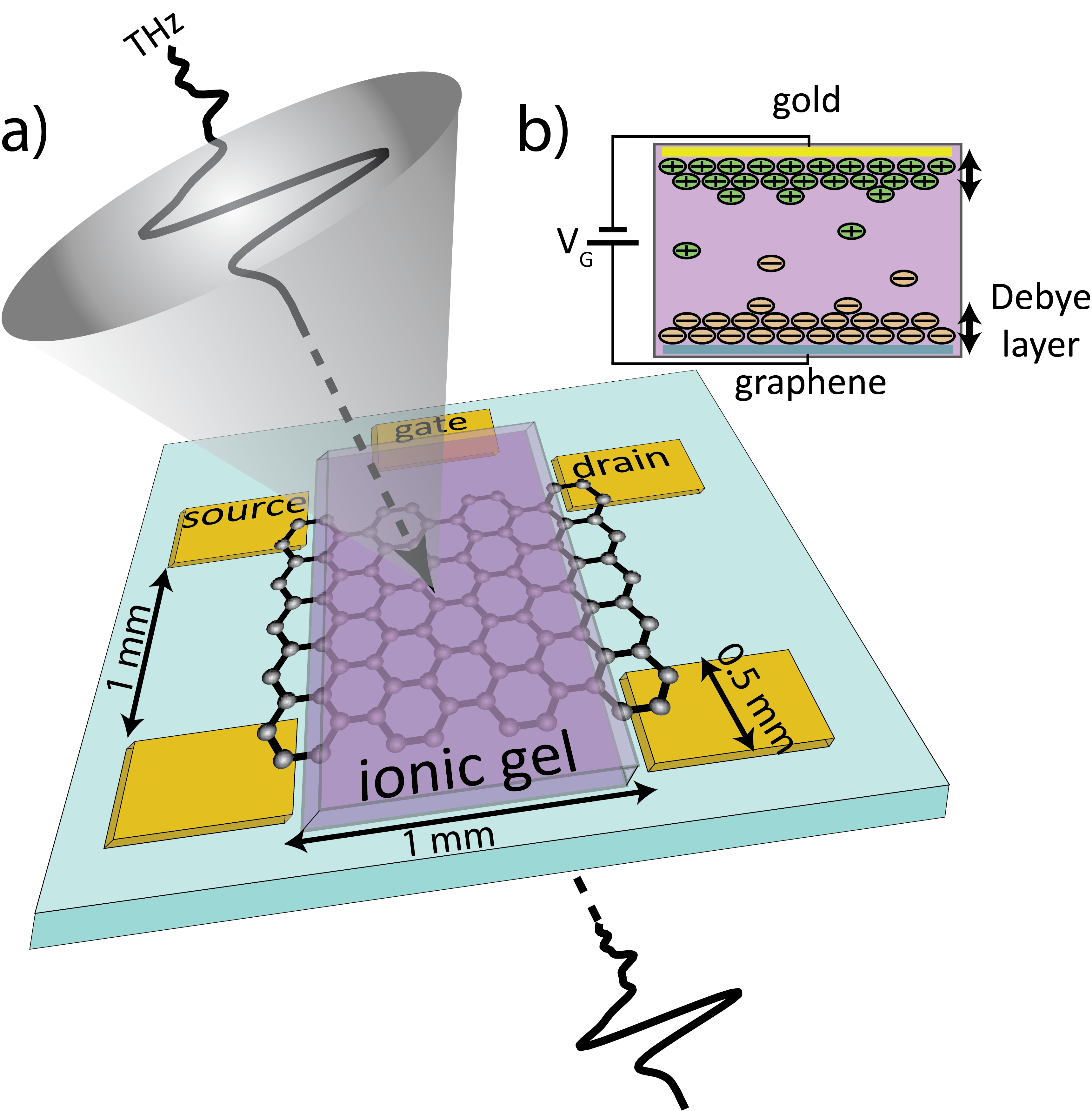}
    \caption{(a) Schematic of the graphene sample on a substrate with gold four-point-probe contacts and a side gate contact. The broadband THz pulses traverse the sample at normal incidence. (b) The mechanism for ionic gating of graphene is shown in the inset, illustrating the accumulated Debye layer under a gate voltage V$_G$. }
    \label{fig:1}
\end{figure}

The well known linear electronic dispersion of graphene naturally gives rise to unusual optoelectronic properties, \cite{novoselov2005two} potentially useful for high-speed modulators and the other active components. While it has been touted as an important material for terahertz (THz) frequency devices, \cite{otsuji2012graphene}
charge transport properties at THz frequencies under field conditions commonly found in devices, on the order 10's kV/cm, are poorly understood. This is
in part due to a lack of control over the extrinsic properties of the devices and graphene's sensitivity to its environment.\cite{docherty2012extreme} The electrodynamic response of doped graphene has previously been studied in the low field, linear regime under gated conditions \cite{wang2008gate} for frequencies ranging from near dc \cite{lin2008operation} to THz,  \cite{dawlaty2008measurement,buron2012graphene,tomaino2011terahertz,shi2014controlling,frenzel2014semiconducting} mid IR \cite{horng2011drude} and optical regimes.\cite{mak2008measurement} Under intense fields, however, charge carriers can acquire significant momentum and for Fermi energies ($E_f$) close to the charge neutrality point (CNP), a complicated interplay between intra- and inter-band transitions can occur.\cite{al2014high} A better understanding of these nonlinear
interactions and their density dependence is needed for the future device design, which is the aim of this study. For this purpose, we study the THz electrodynamic response of graphene under high field conditions and controlled Fermi level through a gate voltage $V_G$ applied to an ionic gel top gate which is transmissive to THz light (see Fig.~\ref{fig:1}).

The development of efficient nonlinear optical methods for the generation of intense THz pulses \cite{hirori2011single} has made it possible to time resolve the high field response of charge and spin excitations in materials.\cite{pashkin2013electric} In graphene, due to a strong interband dipole matrix element that diverges at the charge neutrality (Dirac) point, the THz light-matter interaction can be easily pushed into the nonperturbative regime even by modest electric fields on the order of several 10's kV/cm.\cite{bowlan2014ultrafast,crosse2014theory} As the thermalization time of accelerated hot carriers is noted to be
exceedingly short in graphene, \cite{johannsen2013direct} these nonlinearities can be quantified as an elevated temperature of the electronic subsystem.\cite{MicsNatComm2015}
Recent works have demonstrated a strong THz field-induced transparency in room-temperature graphene \cite{hwang2013nonlinear} in both doped \cite{hwang2013nonlinear,paul2013high,hafez2014nonlinear,paul2014terahertz} and optically excited graphene.\cite{hafez2015nonlinear} This phenomenon has been described by ultrafast intra-band carrier thermalization on the order of 30 fs, \cite{johannsen2013direct} leading to a transient elevated carrier temperature on the order of 2000 K and subsequent optical phonon emission during cooling. In this model, the non-equilibrium heating and generation of optical phonons subsequently enhances the carrier scattering, suppressing the field-driven current and thus THz absorption in graphene. Other sources of carrier scattering that can potentially play an important role in the THz response of graphene include long-range impurity scattering, due to the presence of the charged impurities in the substrate, short-range scattering, due to neutral impurities or disorder in the graphene itself, and carrier-carrier scattering. While impurity scattering dominates at low lattice temperatures, optical phonon scattering becomes more dominant at higher temperatures ($T>$ 700 K).\cite{bao2009nonlinear}

Saturable absorption due to interband contributions has also been observed in the visible regime due to the phase-space filling effects as well as carrier heating. At THz frequencies, the threshold for an interband response is determined by the carrier density, as the carriers must be excited at twice the quasi-Fermi energy to undergo a transition. In this work, we study the Fermi energy dependence of the nonlinear THz transmission of gated graphene and show that although the nonlinear response is similar in different samples with different doping levels, it does not appear that this effect can be described by heating of the lattice alone. The paper is organized as follows. In Sec. II, we discuss the possible scattering mechanisms and their effects on THz transmission as a basis for the model used to describe the data. The results for both five-layer and single-layer graphene samples are presented in Sec. III. Finally, the conclusions are summarized in Sec. IV.

\section{Theoretical model}
In this section, we discuss the model that we use to account for the effects of tuning the Fermi energy on the transmission of THz light. Using this model, we extract the dependence of the scattering time on the Fermi energy and the lattice and electron temperature. There are four potentially important sources of carrier scattering in graphene: short-range impurity scattering, long-range impurity scattering, phonon scattering, and carrier-carrier scattering. 

Strictly speaking the carrier dynamics should be modeled using a full dynamic model that includes the nonlinear response arising from the linear band dispersion and interband transitions. \cite{al2014high} We find however, in agreement with our previous work, \cite{hafez2015nonlinear} that for the samples and field strengths considered here, there is essentially no intrinsically nonlinear response and a linear Drude model is sufficient.  The nonlinear response of the graphene is incorporated in the model by employing a scattering time that depends on the THz field amplitude. Thus, in what follows, we employ the Drude model. 

First, we need to relate the transmission level to the conductivity of graphene. It can be easily shown that the transmission through the graphene sample normalized to the transmitted field through the substrate is given by~\cite{dawlaty2008measurement}
\begin{equation}
T(\omega)=\frac{T_{G+S}}{T_{S}}=\frac{1+n}{1+n+Z_{o}\sigma(\omega) N},
\label{Eq:trans}
\end{equation}
where $T_{G+S}$ and $T_{S}$ are the transmission through the sample and the substrate only, respectively, $n$ is the refractive index of the substrate, $\sigma(\omega)$ is the complex ac sheet conductivity of a single layer of graphene and $N$ is the number of graphene layers
in the sample. The conductivity, dominated by the intra-band response at THz frequencies is given by a simple Drude model ~\cite{horng2011drude}

\begin{equation}
\sigma(\omega)=\frac{D}{(1/\tau-i\omega)},
\label{Eq:sigma}
\end{equation}
where $\omega$ is the angular frequency, $\tau$ is the total carrier momentum scattering time, and $D$ is the Drude weight given by\cite{frenzel2014semiconducting, shi2014controlling}

\begin{equation}
D=\frac{2e^{2}k_{B}T_{e}}{(\pi\hbar^{2})}ln\left[2cosh\frac{{\mu_{f}}}{2k_{B}T_{e}}\right],
\label{Eq:D_weight}
\end{equation}
where $e$ is the electron charge, $k_B$ is the Boltzmann constant, $T_e$ is the electron temperature, $\hbar$ is the reduced Planck's constant, and $\mu_f$ is the chemical potential. For simplicity, the chemical potential is considered to be equal to the Fermi energy at room temperature. However, as the electron temperature rises, we adjust the chemical potential so that the net charge on the graphene is unchanged as the temperature rises. The adjusted chemical potential along with the corresponding temperature are used then to calculate the Drude weight. Using Eqs.~(\ref{Eq:trans}-\ref{Eq:D_weight}), we can extract the total effective scattering time $\tau$ for any measured transmission level for each gate voltage and THz field amplitude. The gate voltage and field amplitude dependence of the scattering time can therefore be directly obtained from the THz transmission data once the electron temperature is determined.

We now turn to the model that we employ for the scattering time used in the Drude model.  We present a simple model of the effects of the lattice temperature and the carrier density on the scattering times based on three scattering mechanisms: short-range neutral impurity scattering, charged long-range impurity scattering and scattering due to absorption of optical phonons.  Although there can be contributions to the scattering rate due to interactions with acoustic phonons, these are usually considered to be relatively small at the temperatures considered here.\cite{fang2011high}  We do not include carrier-carrier scattering due to the difficulties in accurately modelling its effects and because, for similar Fermi energies and THz field amplitudes, it has been shown that carrier-carrier scattering can be neglected without qualitatively changing the results.\cite{al2015optimizing} We will return to this issue when discussing our field-dependent results.

The expression for the total scattering rate is thus given by
\begin{equation}
\frac{1}{\tau} = \frac{1}{\tau_{imp}}+\frac{1}{\tau_{op}},
\label{Eq:tau}
\end{equation}
where $\tau_{imp}$ is the impurity scattering time  $\tau_{op}$ is the optical phonon scattering time.  The impurity scattering rate is given by
\begin{equation}
\frac{1}{\tau_{imp}}=\frac{1}{\tau_{sr}}+\frac{1}{\tau_{Coul}}+\frac{1}{\tau_{ml}}.
\end{equation}
In this expression, ${1}/{\tau_{sr}}$ is the short-range neutral impurity scattering rate, which has been shown theoretically to be linearly dependent on the energy of the carriers, \cite{hwang2008single}; in our model, we take this rate to be proportional to the average energy per carrier with respect to the Dirac point $\textbf{K}$, which is given by
\begin{equation}
E_{av}=\frac{\underset{\mathbf{k}}{\sum}\left\{ \rho_{ee}(\mathbf{K}+\mathbf{k})+\rho_{hh}(\mathbf{K}+\mathbf{k})\right\} \hbar v_{F}\mathbf{k}}{\underset{\mathbf{k}}{\sum}\left\{ \rho_{ee}(\mathbf{K}+\mathbf{k})+\rho_{hh}(\mathbf{K}+\mathbf{k})\right\} },
\end{equation}
where $\rho_{ee}(\rho_{hh})$ is the free electron (hole) population density, and $v_{F}$ is the Fermi velocity.  The second term in the impurity scattering rate is, $\tau_{Coul}$, which is due to charged long-range impurities.  This rate has been shown to be inversely proportional to the square root of the carrier density (due to carrier screening)\cite{dassarmaRevModPhys2011}.  Finally, $\tau_{ml}$ is an additional gate-voltage-independent scattering time that we find is necessary in order to model the bias dependence of the scattering time in our multi-layer sample.  The physical origin of this term is not clear, but it may be due to scattering from the ionic gel or is perhaps arises due to a correction term in the relationship between the gate voltage and the Fermi energy. Thus, the final expression for the impurity scattering rate is
\begin{equation}
\frac{1}{\tau_{imp}} = c_{sr}E_{av}+\frac{c_{Coul}}{E_{eff}}+\frac{1}{\tau_{ml}},
\label{Eq:ci}
\end{equation}
where $c_{sr}$ and $c_{Coul}$ are constants that are experimentally determined and are dependent on the impurity densities, and $E_{eff}$ is the effective Fermi energy that corresponds to the square root of the carrier density and is given by $D\pi\hbar^2/e^2$.  

We now turn to the scattering due to optical phonons.  There are two potential types of optical phonons that can contribute to scattering: optical phonons in the graphene itself and polar surface phonons in the substrate.\cite{berciaud2010electron,konar2010effect} We find, using the parameters in Ref.~\citen{konar2010effect} that, due to the weak coupling strength, the scattering rate contribution from the surface phonons in both substrate materials is much smaller than those due to the graphene optical phonons. Thus, in what follows, we neglect the phonons in the substrate and include only the contribution from the phonons in the graphene itself.\cite{fang2011high}

Previous studies have shown that energy transfer from electrons to phonons takes place on a timescale of about 100 fs.\cite{breusing2009ultrafast,lui2010ultrafast} Thus, in principle, if there are electrons that have been driven by the field more than an optical phonon energy above any vacant states, one would expect emission of optical phonons over the pulse duration, which would increase the lattice temperature.  However, without performing a detailed simulation that tracks the energies of individual electrons, it is very difficult to accurately include this contribution to the scattering.  Thus, for simplicity, we only include the process of phonon absorption, but not phonon emission. 

The optical phonon scattering rate due to optical phonon emission in the graphene is given by
\begin{equation}
\frac{1}{\tau_{op}}=\frac{D_{o}^{2}}{(2\hbar^{2}v_{F}^{2}\sigma_{m}\omega_{o})}\frac{(E_{av}+\hbar\omega_{o})}{(e^{\hbar\omega_{o}/k_{B}T_{l}}-1)},
\label{Eq:tau_op}
\end{equation}
where $\hbar\omega_{o}=147$ meV is the graphene optical phonon energy, $\sigma_{m}=7.6\times10^{-8}$ g/cm$^{2}$ is the 2D mass density of graphene \cite{fang2011high}, $D_{o}=5\times10^{9}$ eV/cm is the deformation potential for the optical phonons\cite{sule2012phonon} and $T_{l}$ is the lattice temperature. Various values for the optical phonon energy and deformation potential have been reported in the literature \cite{strait2011very,sule2012phonon,hwang2009screening,fang2011high}. We employ those found in Ref.~\citen{hafez2015nonlinear}, although, as we discuss in the following sections, we have considered the effect of modifying the deformation potential on the results. We finally note that in all cases we assume equality of the carrier and lattice temperatures that appear respectively in the Drude weight (Eq.~(\ref{Eq:D_weight})) and the optical phonon scattering rate (Eq.~(\ref{Eq:tau_op})).

\section{Experimental Results and Discussion}
In order to investigate the effect of the substrates and number of graphene layers on the transport properties of graphene to the intense THz field, we fabricated two different samples treated in the next two sections. The first is a five layer graphene sample, fabricated by transferring sequentially monolayer graphene onto a sapphire substrate, and the second a monolayer sample on a quartz substrate. The substrates were patterned with $5$/$100~$nm thickness of Cr/Au (prior to transfer) for the electrical four-point-probe measurements as shown in Fig.~\ref{fig:1}. The gate was created by placing the fifth contact at the edge of the substrate so that it is in contact with the ionic gel for electrochemical doping. The ionic gel was fabricated using the method proposed by Ref. \citen{chen2011controlling} and transferred to the sample under an inert glovebox environment. A uniform layer of ionic gel with a thickness of less than $50~\mu$m was spin coated on the graphene sample as well as the bare substrate to serve as a reference. Upon applying the gate voltage, mobile ions are transferred to the surface of the graphene to form a Debye layer with a thickness on the order of $1$~nm.\cite{das2008monitoring} This type of solid polymer electrolyte gate is much thinner than the conventional $300$~nm SiO$_{2}$ back gate, thus a much higher gate capacitance can be achieved. With the electrolyte ionic gating, high doping levels on the order of $5\times10^{13}$~cm$^{-2}$ are possible.\cite{das2008monitoring} The gating response of the graphene is measured in terms of the electrical current driven from the source to the drain electrode through the graphene channel as a function of $V_G$. In order to estimate the Fermi energy as a function of gate voltage relative to the charge neutrality point voltage ($V_{CNP}$), we use the following equation from Ref. \citen{shi2014controlling} for a similar geometry and ionic gel, where
\begin{equation}
E_f=0.346\sqrt{\left(V_{G}-V_{CNP}\right)}~~~[eV].
\label{Eq:Fermi}
\end{equation}
The case of $V_{G} =V_{CNP}$ needs to be treated separately because, in agreement with other researchers, \cite{martin2008observation} we find that due to spatial inhomogeneity of the charge density across the graphene sample an effective $E_f$ must be determined at the nominal CNP gate voltage, in a range of $\pm100$ meV.\cite{shi2014controlling}

The high-field time-domain THz spectrometer was developed based on tilted-pulse front optical rectification in lithium niobate,\cite{hirori2011single} with peak THz field amplitudes limited to 70 kV/cm in this work. The bandwidth of the THz pulses was  $\sim3$ THz and was detected using electro-optic sampling in a $300~\mu$m thick (110)-cut GaP crystal. The emitted THz pulses were collected and focused by a set of off-axis parabolic mirrors to a $450~\mu$m spot size as determined by an uncooled microbolometer camera. The graphene sample
was positioned carefully at the THz focus for high-field transmission measurements at normal incidence. The amplitude of the THz pulses was varied using a pair of wire-grid polarizers prior to the sample in the collimated section of the beam. The measurements were performed by sweeping the graphene gate voltage and registering the value of the transmitted peak THz electric field. No THz pulse reshaping was observed as the gate voltage varies from $-2.5~V$ (below the Dirac point) to $2.5~V$, thus all information on the nonlinear transmission can be quantified by monitoring the THz pulse peak field transmission.

\begin{figure}[t]
    \centering
    \includegraphics[width=6.5 cm]{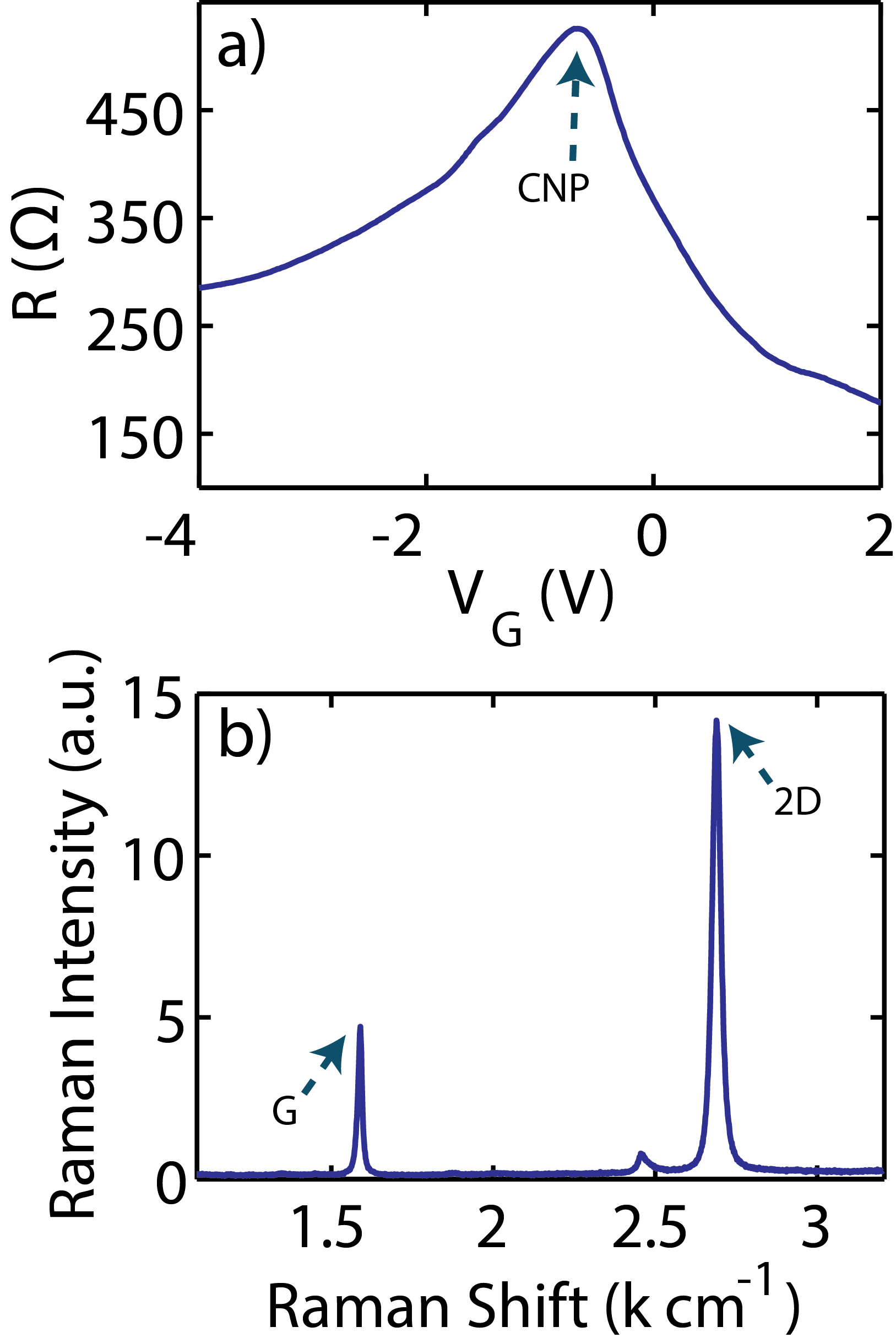}
    \caption{(a) Four-point-probe measurements of the 5-layer sample dc resistance with the charge neutrality point indicated by the resistance peak. (b) The Raman data (with the G and 2D peaks) of single layer of CVD grown graphene before stacking to form the 5 layers.}
    \label{fig:2}
\end{figure}

\subsection{Five-layer sample}
Large-area, single-layer graphene was fabricated by chemical vapor deposition (CVD) on Cu-foil using standard techniques. Poly(methyl methacrylate) (PMMA) is used to transfer the graphene to a c-cut sapphire substrate with refractive index of 3.31 and the PMMA layer was removed. The process was repeated five times so that the five-layer graphene sample was obtained, with the layers electrically decoupled from one another. Fig.~\ref{fig:2}(a) shows the sheet resistance of graphene as a function of gate voltage. The CNP is specified as the resistance maximum which occurs in this case at $V_{G}=-0.8~V$. We note that hysteresis was observed in these resistance measurements when sweeping the gate voltage, which is commonly observed in ionic gel gates.\cite{lee2012all,ye2011accessing} We therefore only perform measurements on the positive sweep of the gate voltage for repeatability. The Debye absorption of the ionic gel is located at lower frequencies than THz, producing a transparent top gated graphene sample. All measurements in this work are performed at room temperature. Fig.~\ref{fig:2}(b) shows the Raman spectroscopy of one of the five layers of graphene before stacking. The ratio between the 2D and G peak values is 2.8 which indicates high quality single layer graphene.\cite{graf2007spatially}

\begin{figure}[t]
    \centering
    \includegraphics[width=8 cm]{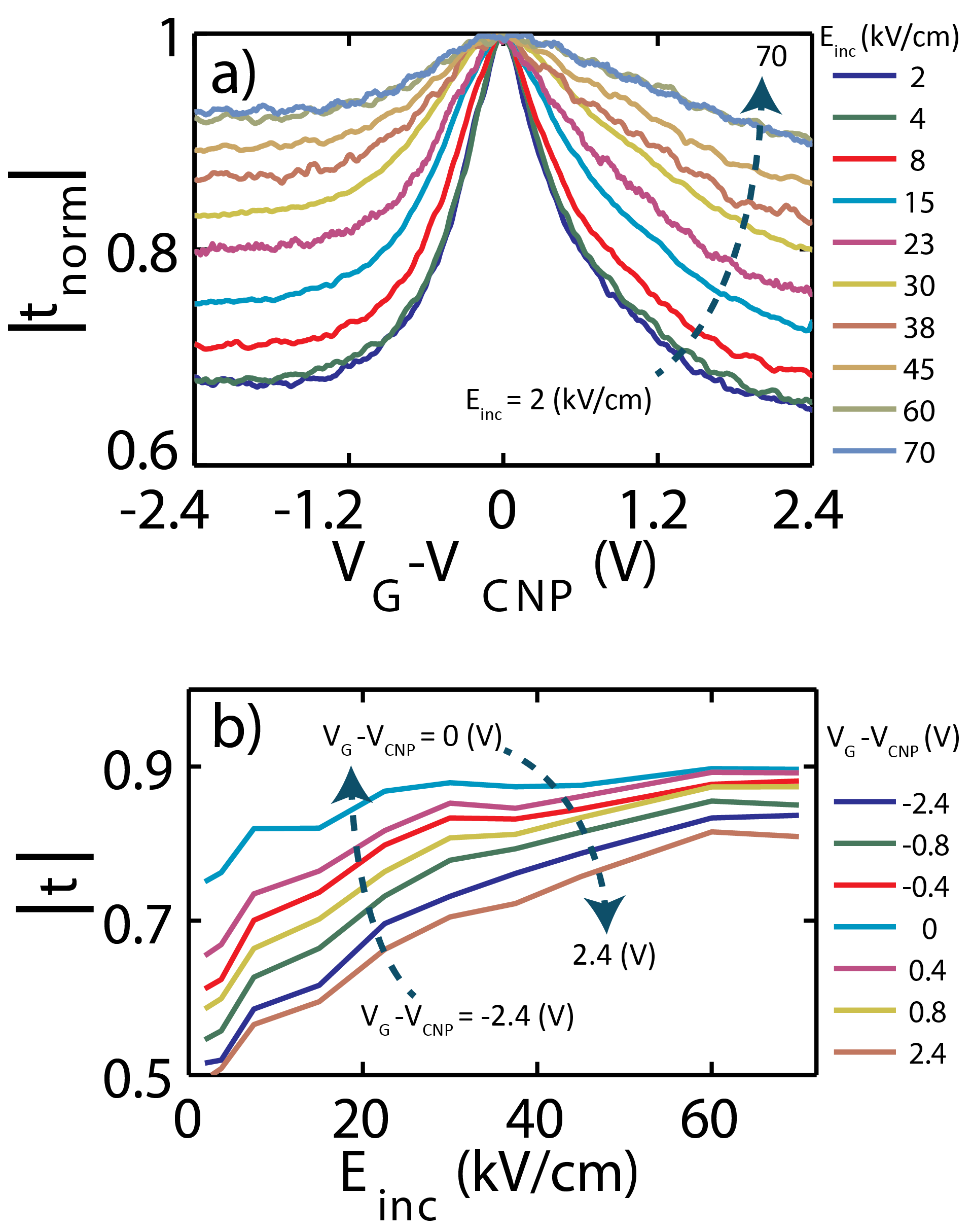}
    \caption{For the five-layer sample: (a) The THz pulse peak field transmission through the gated graphene sample for the field strengths indicated, normalized to the transmission at the Dirac point at  $V_{G}-V_{CNP}=0$. (b) The field dependence of the THz peak field transmission $\mid t \mid =\frac{E_{G+S}}{ E_{S}}$ at various gate voltages. $E_{inc}$ is the incident THz electric field at the graphene sample position.}
    \label{fig:3}
\end{figure}

\begin{figure}[t]
    \centering
    \includegraphics[width=8 cm]{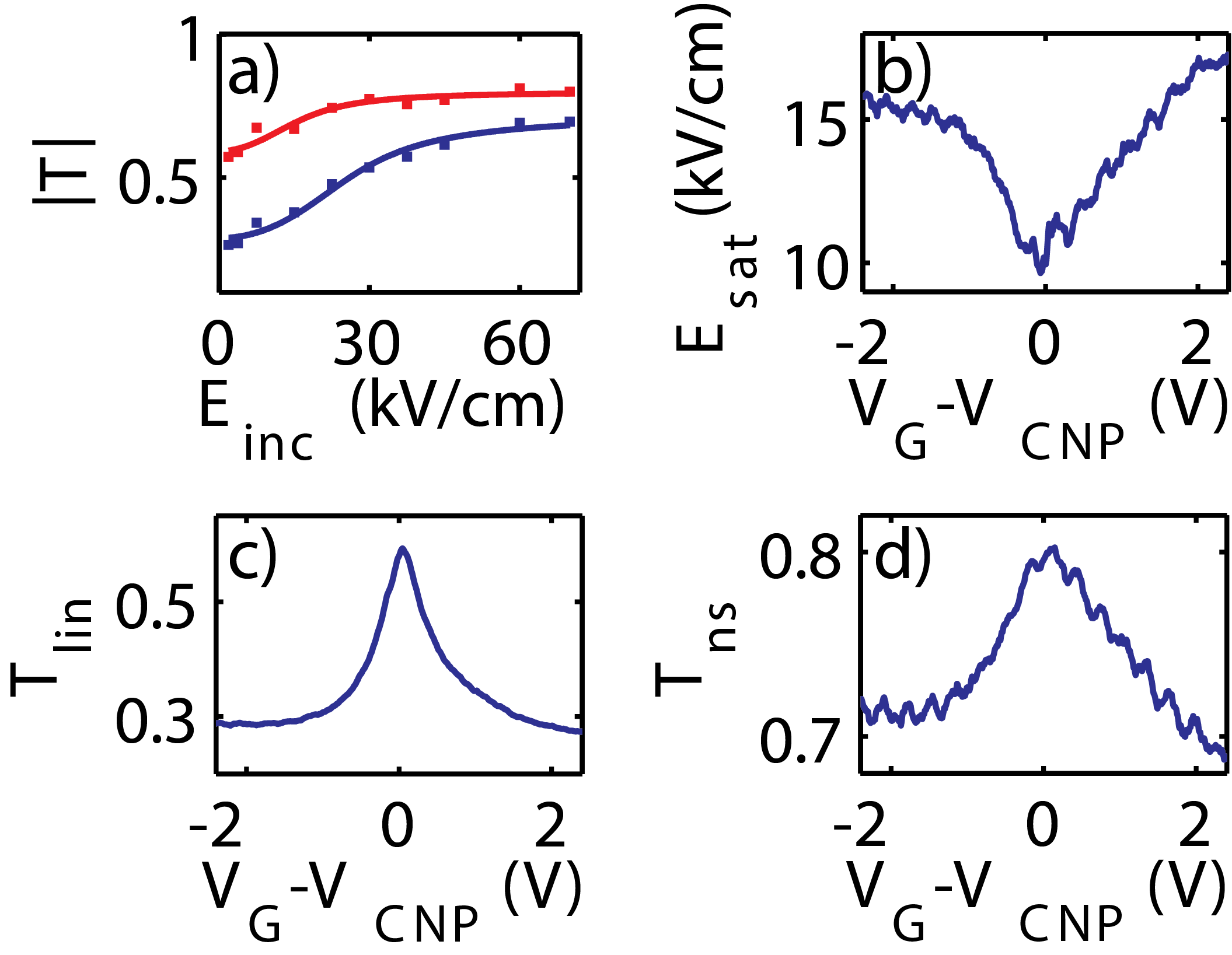}
    \caption{For the five layer sample: (a) THz power transmission defined as $\mid T \mid =\frac{ E_{samp}^2}{ E_{ref}^{2}}$, curve
    fitted by Eq. \ref{saturablepowertransmission} for two different gate voltages, one at the Dirac point ($V_{G}-V_{CNP}=0~V$, red line) and the other one at a n-doped Fermi level corresponding to ($V_{G}-V_{CNP}=-2.4~V$, blue line). (b-d) Fit parameters $E_{sat}$, $T_{lin}$, and $T_{ns}$ extracted from
    Eq. (\ref{saturablepowertransmission}) for power transmission curves as a function of gate voltage.}
    \label{fig:4}
\end{figure}

Fig.~\ref{fig:3}(a) shows the variation of the transmitted peak THz electric field (normalized to its maximum value at the Dirac point) as a function of gate voltage sweep for different incident peak electric fields in the range of $2~$kV/cm to $70~$kV/cm. The THz modulation induced by the gate voltage sweep (Fermi level change in graphene) is a strongly nonlinear function of the peak electric field value. At low fields, gate-induced THz modulation is more than $31 \%$ whereas for higher field amplitudes the modulation decreases and saturates at a value of $7.4$\% above 60 kV/cm. As in previous work on ungated graphene, a saturable power transmission function can be used to phenomenologically describe the behavior observed in Fig.~\ref{fig:3}(b).\cite{hwang2013nonlinear}  The power transmission is defined as $\mid T \mid = \left(\frac{ E_{samp}}{ E_{ref}}\right)^{2}$, where $E_{samp}$ is the peak THz electric field transmitted through the ionic gel, 5 layers of graphene and the substrate, and $E_{ref}$ is the peak field transmitted through the ionic gel and the substrate only. The model is given by:\cite{hwang2013nonlinear}

\begin{equation}\label{saturablepowertransmission}
T({E_{inc}}) = {T_{ns}}\frac{{\ln [1 + \frac{{{T_{lin}}}}{{{T_{ns}}}}({e^{{E_{inc}}^2/{E_{sat}}^2}} - 1)]}}{{{E_{inc}}^2/{E_{sat}}^2}}
\end{equation}
where $E_{inc}$, $E_{sat}$, $T_{lin}$ and $T_{ns}$ are the incident peak electric field, the saturation electric field, the linear power transmission coefficient and the nonsaturable power transmission coefficient, respectively.

In Fig.~\ref{fig:4}(a) we present the measured transmitted power as a function of the incident field along with the fit using Eq.~(\ref{saturablepowertransmission}) for $V_{G} =V_{CNP} $ (corresponding to the CNP, red line) and $V_{G}-V_{CNP}=-2.4~V$ (i.e. highly the n-doped case, blue line). The saturation onset of the power transmission, $\vert T\vert$, occurs at a lower E$_{inc} = 9.7$~kV/cm at the CNP relative to what occurs for higher doping, as demonstrated by Fig.~\ref{fig:4}(b). $T_{lin}$ [Fig.~\ref{fig:4}(c)] and $T_{ns}$ [Fig.~\ref{fig:4}(d)] are higher at the CNP than at other Fermi levels, overall expected due to the larger conductivity of the doped graphene. The saturation behavior for the curves in Fig.~\ref{fig:4}(a) is explained by a decrease in carrier mobility as the THz excitation redistributes carriers within the conduction band.\cite{hwang2013nonlinear} Therefore, the decrease in $E_{sat}$ when the gate voltage approaches the Dirac point can be attributed to an increase in the scattering rate when the Fermi level is tuned to the CNP, as we shall discuss shortly. 

\begin{figure}[t]
\includegraphics[width=8 cm]{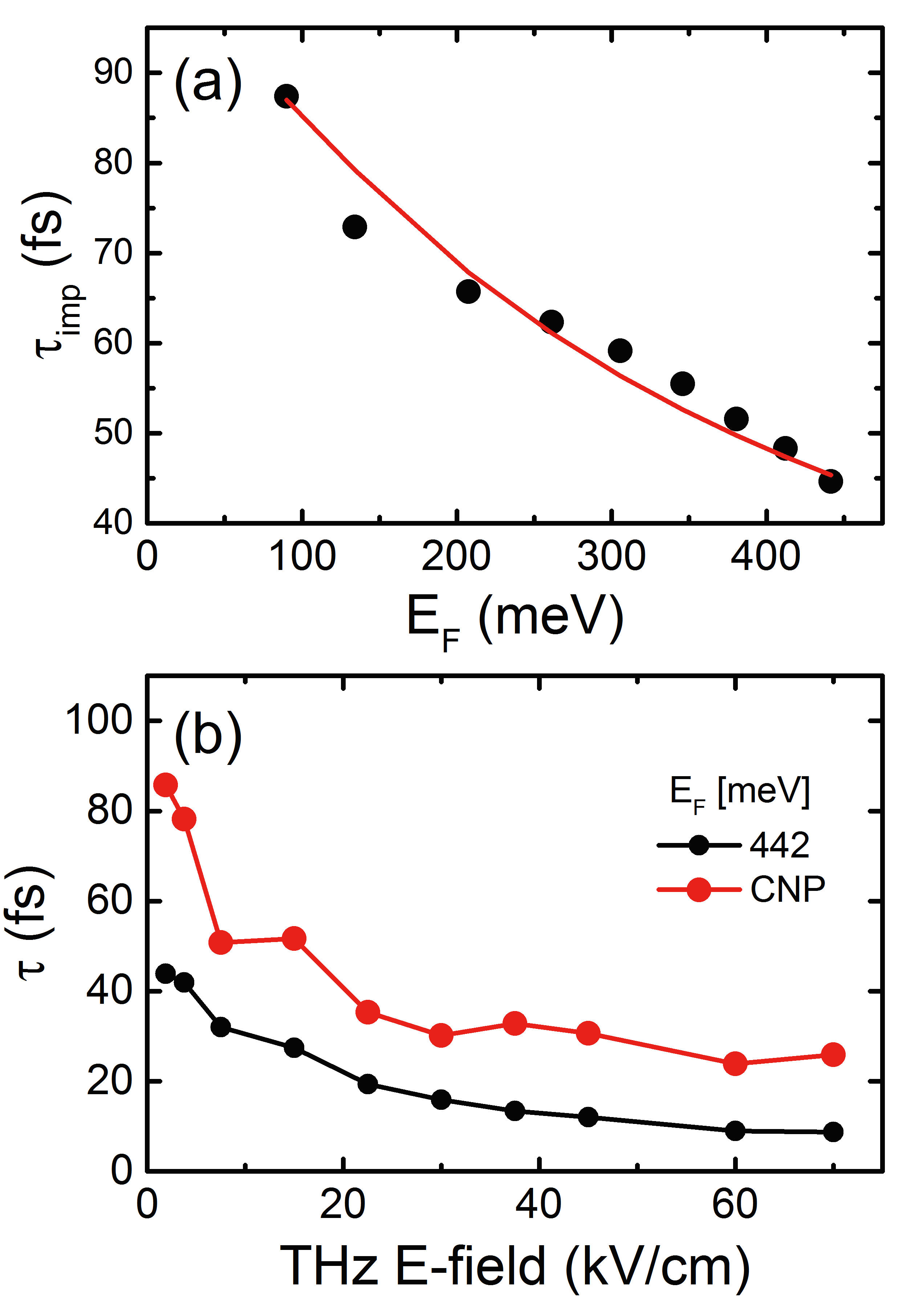}
\caption{(a) The impurity scattering time ($\tau_{imp}$) versus Fermi energy, and (b) total scattering time versus incident field amplitude at the CNP and at the highest Fermi level of 442 meV for the five-layer sample. Note that in (a), the value for $E_f$ the first point is determined via a fit to the curve.}
    \label{fig:5}
\end{figure}

We now use the model presented in Section II to extract the scattering time as a function of the Fermi energy and the incident THz field. While the Fermi energy may not be precisely the same in all 5 layers of the graphene sample, for simplicity we use the same value for the five layers. Similarly, we assume that the scattering time is the same in all layers. We estimate $E_f$ from gate voltages other than the CNP voltage using Eq. \ref{Eq:Fermi}. To determine the Fermi energy for $V_{G} =V_{CNP} $, we first use the model of Section II to extract the fitting parameters, $c_{sr}$, $c_{Coul}$ and $\tau_{ml}$ in Eq. \ref{Eq:ci} for the transmission data for the lowest THz field amplitude of 2 kV/cm. Taking the lattice and electron temperatures at this lowest field to be 300 K, we obtain  $c_{sr}$=49~ps$^{-1} eV^{-1}$, $c_{Coul}=0$ and $\frac{1}{\tau_{ml}}=$7.3~ps$^{-1}$. The data as well as the fit are shown in Fig. \ref{fig:5}(a).  We now extract an effective Fermi energy at the $V_{G} =V_{CNP} $ of approximately 90 meV by extrapolating the fitting curve.  This value is in good agreement with other studies that used the same mechanism to tune the Fermi energy.\cite{shi2014controlling}  We find from our fit that the contribution of charged impurities is negligible and that at this temperature and field strength, the contribution of the optical phonons is also very small.  Thus, the dominant contribution to scattering at this lowest field appears to arise from short-range impurities, which results in the approximately hyperbolic relationship between $\tau$ and $E_f$ seen in the figure.\cite{hwang2008single}

We now turn to the dependence of the scattering time on the incident field for different Fermi energies.  Rather than using our temperature-dependent model of scattering, we first assume that the electron and phonon temperatures do not change with the incident THz field amplitude and simply extract the net scattering time as a function of the incident field strength to two different Fermi levels.  This is plotted in Fig.~\ref{fig:5}(b). As can be seen, for both Fermi energies, the scattering time decreases dramatically with increasing field amplitude. We see that the decrease in the scattering time with field is much larger for the low Fermi energy case.  This would be expected if the increased scattering is due to a mechanism that has a rate that is roughly proportional to the Fermi energy and that would increase at high fields.  There are two likely candidates: optical phonon scattering and carrier-carrier scattering.  

We first consider optical phonon scattering. When the field is higher, it heats the graphene up considerably; this, in turn results in a larger optical phonon population and a shorter scattering time. To test the viability of the optical phonon mechanism, we employed our model that includes the effects of electron and lattice temperature.  Using the impurity scattering rates found at the lowest field and requiring that the electron and phonon temperatures are equal, we were able to obtain good agreement with the experimental data for the transmission by using the temperature as a fitting parameter.  We found that at the highest fields, the temperature required was approximately 1550 K. To determine if such temperatures are physically sensible, we compared the experimental result for the energy absorbed per unit area to the increase in the electron and phonon populations per unit area. We found that the required increase in the system energy was more than an order of magnitude greater than the absorbed THz energy for the 70 kV/cm field.  Most of the increase in the energy of the system when the temperature increases to such levels resides in the optical phonons, not the electrons.  Because there is not universal agreement as the values of the optical phonon energy and deformation potential, we tried adjusting these parameters.  However, we could not find any values of these parameters that yielded anything approaching energy balance at all fields and Fermi energies for one set of parameters.  We conclude from this, that for this sample at least, the decrease in the scattering time with increasing field does not arise entirely from an increase in the optical phonon population due to an increase in the lattice temperature. 

Another possible mechanism for the field dependent scattering is carrier-carrier scattering. This scattering rate increase rapidly as the carrier density increases (as observed).  Moreover, it also increases with the energy of the carriers, which will occur when the field amplitude is high. Modeling carrier-carrier scattering with energy dependence would require a sophisticated Monte-Carlo or density matrix simulation. Thus, although we cannot rule out this mechanism, it is beyond the scope of the current work.

We now examine the results for our monolayer graphene sample to see how its transmission depends on the Fermi energy and the THz field amplitude.

\begin{figure}[t]
\includegraphics[width=6.5 cm]{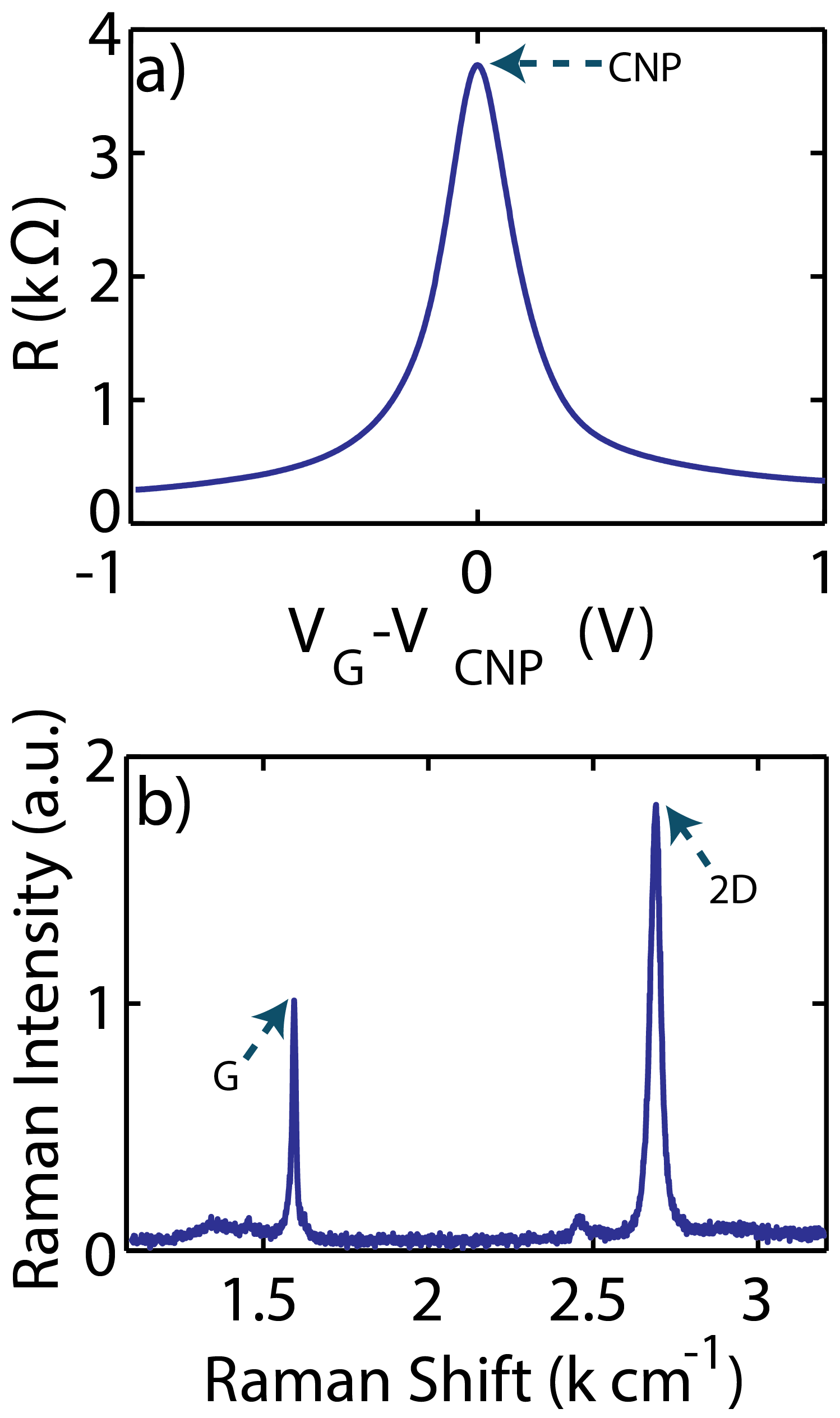}\caption{(a) Two terminal measurements of the resistance and (b) the Raman data (with the G and 2D peaks) of the monolayer sample.}
    \label{fig:6}
\end{figure}
\begin{figure}[t]
\includegraphics[width=7.5 cm]{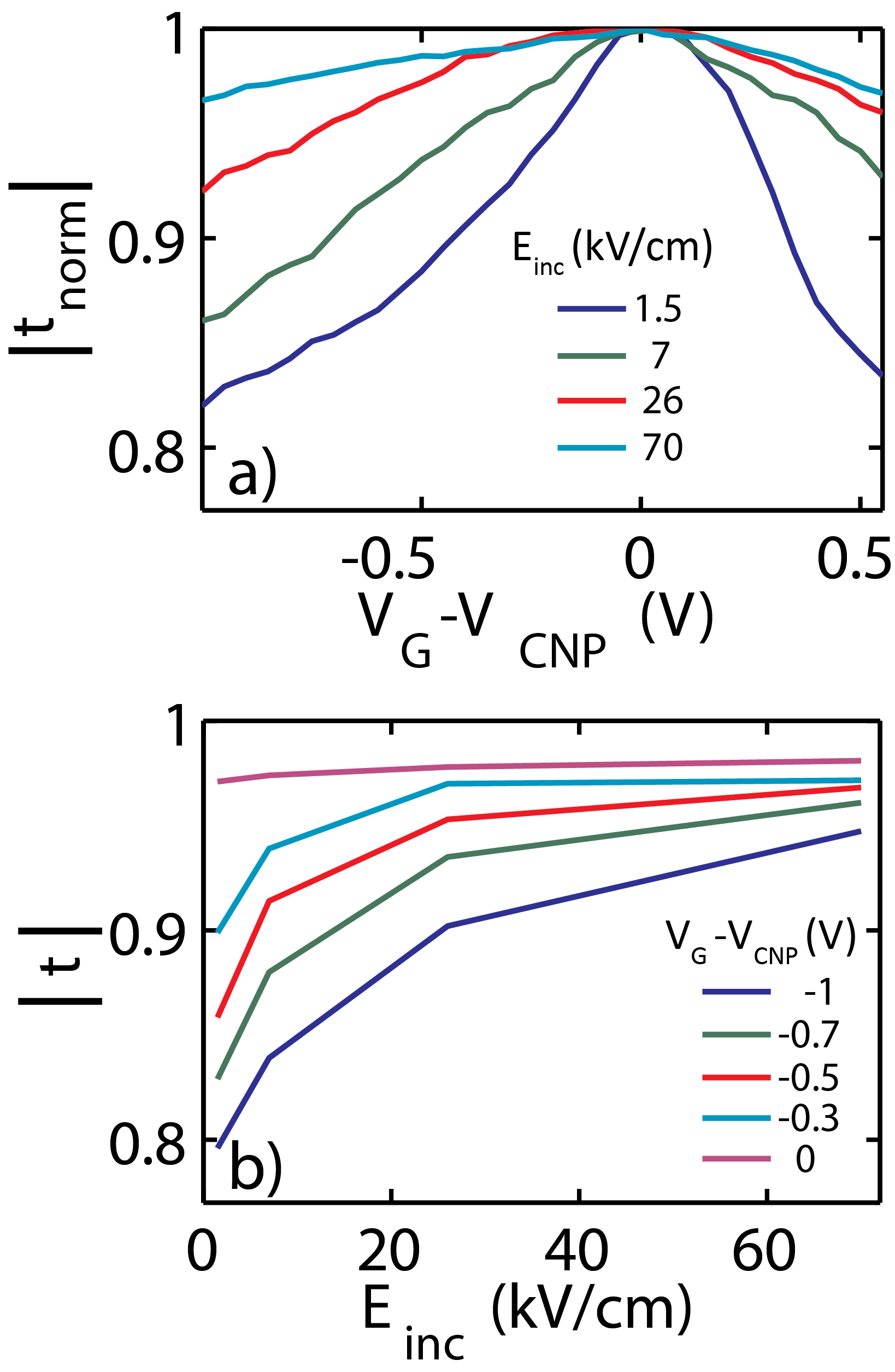}\caption{(a) The normalized peak transmission as a function of the gate voltage for various THz peak electric field strengths and (b) the dependence of the peak transmission on the THz peak electric field for various doping levels for the monolayer sample.}
    \label{fig:7}
\end{figure}
\subsection{Single-layer sample}
A single-layer graphene sample was prepared by chemical vapor deposition on Cu-foil and transferred onto a z-cut quartz substrate with a refractive index $n= 1.96$. The sheet resistance of this sample as a function of $V_G$ was determined by a two terminal measurement shown in Fig. \ref{fig:6}(a) with the CNP indicated again by the resistance peak. Electron (hole) doping is induced by changing $V_G$ above (below) the CNP voltage $V_{CNP}$ and Eq. \ref{Eq:Fermi} is used to calculate the Fermi level energy. Fig.~\ref{fig:6}(b) shows the Raman spectra for the single layer graphene sample, with a 2D/G peak ratio of 1.78 confirming high quality monolayer graphene.

Fig. \ref{fig:7}(a) shows the normalized THz peak electric field amplitude transmitted through the graphene when the Fermi-level is swept from -1 to 0.5 V. Again, our data does not show any phase change in the THz waveform transmitted through the graphene by increasing the incident THz field and so the conductivity change can be completely quantified by monitoring the peak transmission. For the lowest THz peak electric field strength of 1.5 kV/cm, a significant 20\% reduction in the peak transmission was observed at V$_G$ - V$_{CNP} = -1$ V or E$_f = -340$ meV. As expected, the increased doping leads to higher conductivity and therefore absorption of the THz pulse, however, increasing the incident THz field leads to a reduction in the doping-induced absorption as in the 5-layer sample.\cite{paul2013high,hafez2015nonlinear} As in the 5-layer case, a continuous reduction in field-induced transmission modulation occurs as the gate voltage approaches the charge neutrality point. This behavior is best seen in Fig.~\ref{fig:7}(b), showing the peak transmission as a function of the THz peak field strength for various applied V$_G$. The THz field-induced transparency (the percentage increase in transmission) is more pronounced for high doping levels.  As with the five-layer sample, we find that the induced nonlinearity is mainly due to an increase in the carrier scattering when the THz field is increased.\cite{paul2013high, hafez2015nonlinear} 

\begin{figure}[t]
\includegraphics[width=8 cm]{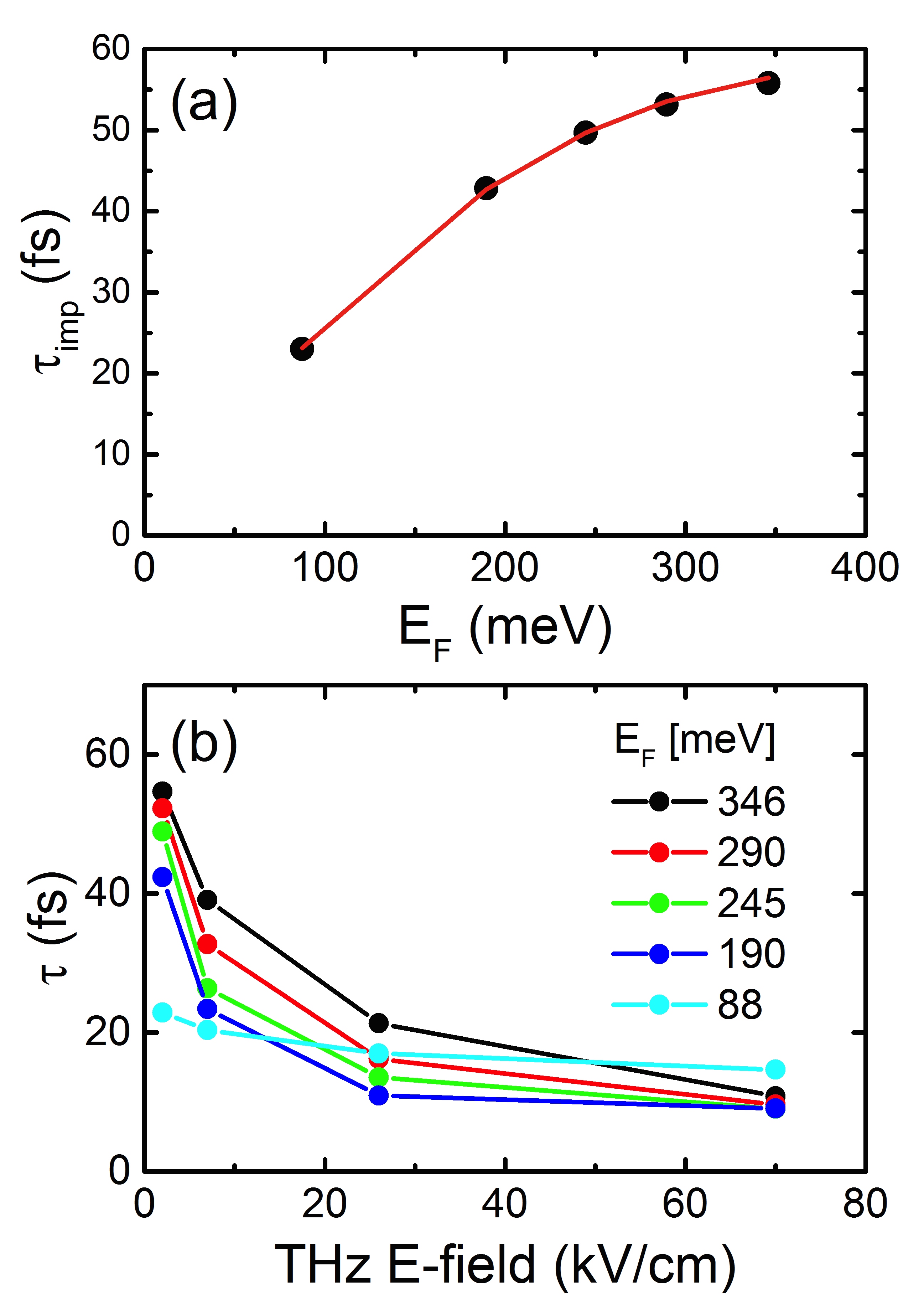}\caption{(a) The impurity scattering time ($\tau_{imp}$) versus Fermi energy at the lowest incident THz field of 1.5 kV/cm, and (b) the THz field dependence of (a) the carrier scattering rate. Note that in (a), the value for $E_f$ the first point is determined via a fit to the curve.}
    \label{fig:8}
\end{figure}

We extract the scattering behavior by using the model of Section II.  We extract the low-field scattering parameters in the same way as was done in the last section taking the temperature to be room temperature. The transmission data at the lowest field amplitude can be well reproduced by the theory only if we consider both short-range neutral and long-range charged impurity scattering in addition to optical phonon scattering. We find that the best fit is obtained for $c_{sr}=$30~$ps^{-1}eV^{-1}$, $c_{Coul}=$3.63 $ps^{-1}eV$ and and $\tau_{ml}=$0. The data as well as the fit are shown in Fig. \ref{fig:8}(a).  Note that the increase in the scattering time with Fermi energy indicates that long range charged impurity scattering is dominant at this field.  The presence of long range scattering in this sample may be linked with the direct contact of the graphene with the SiO$_2$ substrate as opposed to the 5 layer sample, where 4 of the layers are physically removed from the substrate and the effects of charged impurities on carriers in the upper layers are partially screened by carriers in the lower ones. Finally, using the extrapolation of the fit to low Fermi energy, we extract and Fermi energy of 88 meV for the bias V$_G$ = V$_{CNP}$.  


We now use the same constant-temperature model of the previous section to determine the net scattering time as a function of the incident field strength for the five different Fermi levels and plot these times in Fig.~\ref{fig:8}(b). As in the five-layer case,  the scattering time decreases dramatically with increasing field amplitude for all Fermi energies.  However, in contrast to the results of the previous section, we find that the decrease in the scattering time with the field is much smaller when V$_G$ = V$_{CNP}$.  One reason for this difference is that in the monolayer, the low-field scattering time is the shortest when the Fermi energy is the lowest, thus the effect of any additional scattering channels is smaller.  At the highest fields, just as in the five-layer sample, the scattering time is the largest when V$_G$ = V$_{CNP}$. This seems to indicate again that the field-dependent scattering rate is larger when the carrier density (or perhaps average carrier energy) is larger. However, the trend is not the same for the intermediate Fermi energies, where at the highest field, the scattering rate is larger when the Fermi energy is smaller (with the exception of the CNP result). It appears that the result at the CNP is somewhat of an anomaly in this sample.

Although, the Fermi energy dependence of the field-dependent scattering component is not as clear as in the five-layer sample, we again examined where optical-phonon scattering is a possible mechanism. We modeled the optical phonon scattering by taking the temperature to be a fitting parameter.  Again, although we could reproduce the observed transmission curves, the temperatures required were very large.  The temperatures at the highest field ranged from 1300 K for the highest Fermi energy to almost 3000 K at the CNP. Again, such huge temperatures require that the energy deposited in the system is a factor of 3 to 10 larger than the energy that is experimentally found to be absorbed by the graphene.  As in the five-layer case, we tried adjusting the phonon energy and deformation potential, but for all combinations that we tried we could  not achieve anything approaching energy balance at all fields and Fermi energies. We thus conclude that for this sample as well, optical phonons cannot be the major contributor to the increase in the scattering rate with increasing field. In this sample, however, due to the more complex dependence of the field-dependent scattering contribution on the Fermi energy, it is not as clear that carrier-carrier scattering is a likely candidate.


\section{Conclusion}
We have studied nonlinear THz light-matter interaction for both monolayer graphene on quartz and a stack of 5 monolayers of graphene on sapphire under the controlled Fermi level conditions.   The two samples exhibited opposite dependencies on the Fermi energy for low THz fields.  The monolayer sample scattering rate was dominated by charged impurities, while the five layer sample scattering was neutral impurity dominated.  This difference is likely explained as arising from different preparation recipes as well as differences in the substrates. 

We also observed a strong increase in the THz transmission with increasing field that occurs for THz peak electric fields as low as 8 kV/cm.  This increase appears to result from a decrease in the carrier scattering times with increasing field. Our analyses indicate that optical phonon absorption is not likely the main source of this field-induced increase in the scattering rate.  The source of this increase remains uncertain.  For the five layer sample, the most promising mechanism is perhaps carrier-carrier scattering, but for the monolayer sample the mechanism is even more unclear.  

This work points to the importance of future modelling of full carrier and lattice dynamics in graphene excited by intense THz pulses to determine the main source of the strongly nonlinear response. The interaction of high field THz pulses with graphene is significantly dependent on Fermi energy. Moreover, given the variability in processing of graphene films, we show that it is entirely possible that different impurity scattering mechanisms dominate in different samples, even though the field dependence is similar for a fixed Fermi energy. Thus any comprehensive study must be under controlled gated conditions. We anticipate that our findings will be important for future designs of graphene devices operating at THz frequencies, influencing performance due to nonlinearities arising at the kV/cm operating fields.

The authors gratefully acknowledge funding support from NSERC, FRQNT and CFI.

\bibliographystyle{apsrev4-1}
\end{document}